\documentclass[twocolumn,english,conference]{IEEEtran}
\ifCLASSINFOpdf
\else
\fi

\usepackage[T1]{fontenc}
\usepackage{babel}
\usepackage{array}
\usepackage{calc}
\usepackage{amsmath}
\usepackage{amssymb}
\usepackage{amsthm}
\usepackage{graphicx}
\usepackage{cite}
\usepackage{subfigure}
\usepackage{color}

\usepackage{enumitem}
\usepackage{multirow}
\usepackage{hhline}
\usepackage[ruled]{algorithm2e}
\usepackage{algpseudocode}

\usepackage[mathlines,switch]{lineno}

\usepackage[unicode=true,
bookmarks=true,bookmarksnumbered=true,bookmarksopen=true,bookmarksopenlevel=1,
breaklinks=false,pdfborder={0 0 0},backref=false,colorlinks=false]
{hyperref}
\hypersetup{pdftitle={Your Title},
	pdfauthor={Your Name},
	pdfpagelayout=OneColumn, pdfnewwindow=true, pdfstartview=XYZ, plainpages=false, bookmarks=false}
\makeatletter

\let\oLRDforeign@language\foreign@language
\DeclareRobustCommand{\foreign@language}[1]{%
	\lowercase{\oLRDforeign@language{#1}}}

\makeatother
\allowdisplaybreaks

\makeatletter
\newcommand{\thickhline}{%
	\noalign {\ifnum 0=`}\fi \hrule height 1.5pt
	\futurelet \reserved@a \@xhline
}
\newcolumntype{"}{@{\hskip\tabcolsep\vrule width 1pt\hskip\tabcolsep}}
\makeatother

\begin{document}
	%
	\title{Unobservable False Data Injection Attacks against PMUs: Feasible Conditions and Multiplicative Attacks}

	
	
	%
	\author{\IEEEauthorblockN{
			Zhigang Chu,
			Jiazi Zhang,
			Oliver Kosut,
			and	Lalitha Sankar}
		\IEEEauthorblockA{School of Electrical, Computer and Energy Engineering\\
			Arizona State University,
			Tempe, AZ, 85287}}


	\maketitle
	\pagestyle{plain}
	\begin{abstract}
		This paper studies false data injection (FDI) attacks against phasor measurement units (PMUs). As compared to the conventional bad data detector (BDD), an enhanced BDD utilizing the effect of zero injection buses is proposed. Feasible conditions under which FDI attacks are unobservable to this enhanced BDD are discussed. In addition, a class of multiplicative FDI attacks that maintain the rank of the PMU measurement matrix is introduced. Simulation results on the IEEE RTS-24-bus system indicate that the these multiplicative unobservable attacks can avoid detection by both the enhanced BDD and a detector based on low-rank decomposition proposed in prior work. 
	\end{abstract}
	

	%
	\IEEEpeerreviewmaketitle
	\global\long\def\figurename{Fig.}
	\global\long\def\tablename{TABLE}

	\section{Introduction}\label{sec:Introduction}
	In the past decade, phasor measurement units (PMUs) have been widely deployed in power systems for monitoring, protection, and control purposes. With the ability to directly measure the bus voltages and phase angles with high sampling rate and accuracy, PMUs have the potential to play a significant role in real-time power system state estimation (SE) \cite{Phadke_PMU1986}, dynamic security assessment \cite{Vittal2007, DSA_PMU2012}, system protection \cite{PMU_application2010}, and system awareness \cite{PMU_monitoring2010}.      

	The communication and computing devices that enable wide-area real-time monitoring and control of the electric power system have been demonstrated to be vulnerable to cyber-attacks \cite{StuxNet2014,UkraineAttack,USAttack2018_2}. As an increasingly important component of these devices, PMUs are also prone to cyber-attacks. Therefore, it is of great importance to evaluate the vulnerability of PMUs to potential cyber-attacks. The authors of \cite{Lin_PMUcybersecurity2012,Beasley2014_PMU} classify the potential cyber-attacks on PMUs as communication link damage attacks, denial of service attacks, data spoofing attacks including GPS spoofing attacks and false data injection (FDI) attacks. 
	
	FDI attacks involve an intelligent attacker who replaces a subset of measurements with counterfeits. 
    We focus on the sub-class of unobservable attacks which render the false data as unobservable to the operator. For supervisory control and data acquisition (SCADA) systems, prior work \cite{Liu2009,Kosut2011,Zhang2016TSG,Liang2015,ZhangPES2016,Chu2016SmartGridComm,Zhang2018,Liang2014} has shown that attacks can be designed to have severe physical \cite{Liang2015,Chu2016SmartGridComm,ZhangPES2016,Zhang2018,Liang2014} and/or economical consequences \cite{Yuan11,Yuan2012}. The authors of \cite{Liang2014} were the first to determine the conditions under which an attacker with limited resources and capabilities can launch an unobservable FDI attack on SCADA measurements.
    However, such conditions are not known for attacks on PMU measurements. Furthermore, the high cost of PMU installation restricts the PMUs to be placed only at a subset of all buses. As a consequence, only a subset of bus voltages and branch currents are directly measured by PMUs. This leads to the new question that we address in this paper: under what conditions is it feasible to launch unobservable attacks against PMUs?

	In \cite{Kim_PMU2013}, Kim and Tong introduce a protection scheme by placing secure PMUs to simultaneously ensure observability and prevent FDI attacks. However, as PMUs are also prone to attacks, their approach cannot thwart FDI attacks when PMU measurements are compromised by attackers. The authors of \cite{StatisticalFIDdetection_2013} propose a decentralized FDI attack detection approach based on the Markov graph of bus voltage angles. The drawback of this approach is that it may not perform well when the system experiences a disturbance. An expectation-maximization based detector is introduced by Lee and Kundur in \cite{Kundur_PMU2014} to detect FDI attacks on PMUs, but it only assumes DC power flow model and requires the bus power injections to be known. 

	%
	%
	Using measurements obtained from deployed PMUs in the grid, \cite{DahalPMU2012} and \cite{ChenPMU2013} illustrate the low-rank nature of PMU data when it is structured in a matrix. Utilizing this low-rank observation, \cite{Wang_SmartGridComm2014_PMU,Gao2016_PMUIdentification} propose a low-rank decomposition (LRD) based detector to detect FDI attacks on PMU measurements.
	On the other hand, the FDI attacks of most interest are those in which the attacker is not omniscient and omnipresent --- this limited knowledge and limited capabilities of FDI attacks are often captured (see, for \textit{e.g.}, \cite{Liu2009,Kosut2011,Liang2015,Zhang2016TSG,ZhangPES2016,Chu2016SmartGridComm,Zhang2018}) by restricting attacker knowledge to a subset of the network and restricting counterfeits to a small number of meters. This latter restriction along with the above mentioned low-rank properties of a block of PMU data suggests that the resulting counterfeit PMU measurement matrix can be viewed as a linear combination of a low-rank (actual) measurement matrix and a sparse attack matrix (counterfeit additions to measurement). 	
	The authors of \cite{Wang_SmartGridComm2014_PMU,Gao2016_PMUIdentification} show that an LRD-based detector can identify the column sparse FDI attack matrix assuming that the attacker attacks the same set of PMU measurements over time. 
	
	In this paper, we propose an enhanced bad data detector (BDD) that utilizes Kirchhoff's current law (KCL) at zero injection buses (ZIBs) in addition to the conventional residual-based BDD. We then identify a sufficient condition under which an attack is unobservable to this enhanced BDD. Furthermore, we propose a class of multiplicative unobservable FDI attacks that multiply the state matrix by a full rank matrix, thereby preserving the rank of the measurement matrix. We illustrate that these attacks can bypass the LRD detector via simulation on the IEEE RTS-24-bus system.
	 
	\section{PMU Measurement Model}\label{sec:PMmodel}
	Throughout, we assume the power system is completely observable by PMUs. A PMU placed at a bus collects the complex voltage measurements at this bus and current measurements on all branches connect to it, typically at a rate of 30 samples per second \cite{Phadke2009_PMUSE}. These measurements are linear functions of the states, \textit{i.e.,} the complex bus voltages. Let $p$ be the number of buses (states), and $n$ be the number of synchrophasor measurements in the power system, at each time instant $t$, the PMU measurement model is given by
	\begin{equation}
	w_t=Hx_t+e_t\label{eq:ACMeasurement}
	\end{equation}
	where $w_t$ is the $n \times 1$ measurement vector; $x_t$ is the $p \times 1$ state vector of complex bus voltages; $e_t$ is an $n\times1$ additive Gaussian noise vector whose entries are assumed to be i.i.d; $H$ is the $n\times p$ measurement Jacobian matrix. The least squares estimate of $x_t$, $\hat{x}_t$, is given by\cite{AburBook}
	\begin{equation}
	\hat{x}_t=H^{+}w_t\label{eq:LinearSE}
	\end{equation}
	where $H^{+}$ is the pseudo-inverse of $H$. The conventional residual-based BDD performs $\chi^2$ test on the residual vector 
	\begin{equation}
		r=w_t-H\hat{x_t}=w_t-HH^+w_t \label{ConvBDD}
	\end{equation}
	to detect bad measurements.
	
	Given an PMU placement scheme, \eqref{eq:ACMeasurement} can be written as 
	\begin{equation}\label{eq:AvaiMeasurements}
	w_t=\begin{bmatrix}
	V_t \\ I_t
	\end{bmatrix} +e_t
	=Hx_t+e_t=\begin{bmatrix}
		\textbf{I}'\\ \textbf{Y}
	\end{bmatrix}x_t+e_t,
	\end{equation}
	where $V_t$ and $I_t$ are available voltage and current measurements at time $t$, respectively; $\textbf{I}'$ is the reduced identity matrix with only rows corresponding to PMU buses; and $\textbf{Y}$ is the dependency matrix between available current measurements and states. Each row of $\textbf{Y}$ corresponds to a current measurement $I_{i,j}$ from bus $i$ and bus $j$. It contains only two non-zero entries, $Y_{ij}$ at the $i$th entry and $-Y_{ij}$ at $j$th entry, where $Y_{ij}$ is the admittance of the line between bus $i$ and bus $j$.
	
	
	PMU data collected over a block of time (\textit{e.g.,} 2 to 20 seconds) can be written as a matrix where each row vector corresponds to all PMU measurements at one time instant and each column vector consists of the measurements collected in the same channel over a period of time. The PMU measurements in \eqref{eq:ACMeasurement} over $N$ time instants can then be collected as
	\begin{equation}
	W = XH^T+ E \label{eq:MeasurementMatrix}
	\end{equation}   
	where matrices $W=\left[w_1^T;\: w_2^T; \hdots ;\:w_N^T\right]$, $X=\left[x_1^T;\: x_2^T;\hdots;x_N^T\right]$, and $E=\left[e_1^T;\: e_2^T;\hdots;e_N^T\right]$ are PMU measurement matrix, state matrix, and noise matrix, respectively. Note that $w_t^T$, $x_t^T$, and $e_t^T$ for $t=1, 2,\hdots,N$ are the transpose of the measurement, state, and noise column vectors, respectively, in \eqref{eq:ACMeasurement}.
	
	
	

	
	\section{Unobservable FDI Attacks on PMU Measurements}\label{sec:uFDI} 
	In an FDI attack, the attacker is assumed to control a subset $\mathcal{J}\subseteq\{1,\ldots,n\}$ of measurements. Thus, the attacker may replace $W$ with
	\begin{equation}
		\bar{W}=W+D, \label{eq:FDIattack}
	\end{equation} where the column support of the measurement attack matrix $D$, $supp(D)$, is contained in $\mathcal{J}$. An attack can bypass the conventional residual-based BDD if $D=CH^T$ for some non-zero matrix $C$, \textit{i.e.,} 
	\begin{equation}
	\bar{W}=W+CH^T. \label{eq:uFDIattack}
	\end{equation}
	From \eqref{eq:LinearSE}, the estimated state matrix without attack is given by
	\begin{equation}
		\hat{X} = W H^{+T},\label{eq:Xhat}
	\end{equation}
	and the estimated state matrix under attack is given by
	\begin{equation}
	    \bar{X} = \bar{W}H^{+T}=WH^{+T}+CH^TH^{+T}=\hat{X}+C.\label{eq:Xbar}
	\end{equation}
	The residual matrix under attack is given by
	\begin{flalign}
		R&=\bar{W}-\bar{X}H^T\notag\\
		&=W+CH^T-(\hat{X}+C)H^T\notag \\
		&=W-\hat{X}H^T\label{eq:R}
	\end{flalign}
	is the same as that without attack. Therefore, attacks in the form of \eqref{eq:uFDIattack} cannot be detected by the conventional BDD.
	\subsection{Enhanced BDD Utilizing ZIBs} \label{EBDD}
\begin{figure}[h]
		\hspace{-0.9cm}\centering \includegraphics[trim=0 0.3cm 0 0.3cm, scale=0.85]{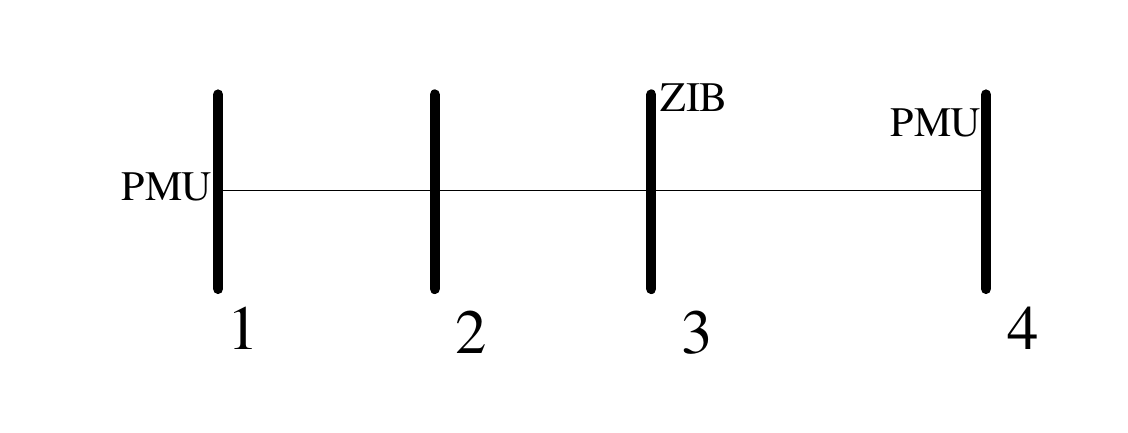}\protect\protect\caption{4-bus system illustrating that FDI attacks on PMUs can be detected in the presence of ZIBs. \label{SmallSys}}
	\end{figure}

	An attack in the form of \eqref{eq:uFDIattack} is perfectly unobservable if the network is comprised of only injection buses. However, if the network includes some ZIBs, \eqref{eq:uFDIattack} is insufficient to keep the attack unobservable. For example, in the 4-bus system shown in Fig. \ref{SmallSys}, assume the attacker controls the PMU at bus $1$ but not that at bus $4$. If the attacker wants to change the state at bus $2$, it needs to change measurement $I_{1,2}$. This attack is in the form of \eqref{eq:uFDIattack}, and hence, can bypass the conventional BDD. However, since the attacker does not control the PMU at bus $4$, it cannot change measurements $V_4$ and $I_{3,4}$. As a result, the state at bus $3$ cannot be changed. If the state at bus $2$ is changed, the current $I_{2,3}$ is changed. Thus, $I_{2,3} \neq I_{3,4}$, so the KCL of ZIB $3$ is violated under this attack.
	
	The example above illustrates the potential of ZIBs in detecting FDI attacks. Let $k$ be the number of ZIBs in the system. For the $i^\text{th}$ ($i \in {1,\ldots,k}$) ZIB $q$, let $\mathcal{M}$ be the set of all its neighboring buses, the dependency matrix $A \in \mathbb{C}^{k\times p}$ between injections at ZIBs and states is given by
		\begin{equation}
		A_{ij} = \left\{ \begin{array}{lr}
		-\sum\limits_{l}Y_{lq}, \hspace{0.5cm}j=q, l \in \mathcal{M} \\
		Y_{jq},  \hspace{1.25cm}j \in \mathcal{M}\\
		0,  \hspace{1.6cm}j \notin \mathcal{M}.
		\end{array}\right.
		\end{equation}
	For instance, in the system shown in Fig. \ref{SmallSys}, $k=1$,
	\begin{equation}
	A = [0, \hspace{0.1cm} Y_{23}, \hspace{0.1cm} -(Y_{23}+Y_{34}), \hspace{0.1cm} Y_{34}].
	\end{equation}
	
	We define a new residual matrix for the enhanced BDD 
	\begin{flalign}\label{eq:Rnew}
		R_e = \begin{bmatrix}
		\bar{W}-\bar{W}H^{+T}H^T\\\bar{X}A^T
		\end{bmatrix}=\begin{bmatrix}
		R\\\bar{X}A^T
		\end{bmatrix}
	\end{flalign}	
	In the absence of noise, an attack-free state matrix $X$ always satisfies $XA^T=0$. The enhanced BDD performs $\chi^2$ test on $R_e$, and bad data is flagged if $R_e$ fails the test.
	
\subsection{Feasible Conditions for Unobservable Attacks}\label{FeasCondition}
	As stated in the Introduction, the cost of placing PMUs in power systems is much higher than that of placing power meters. Because of this cost, PMU measurements are typically available only at a subset of buses, in contrast to SCADA measurements that are available almost everywhere in the network. Besides, PMUs and SCADA system collect different types of measurements. Thus, the conditions for launching unobservable FDI attacks on SCADA system do not directly generalize to PMUs.
	
	An attack is defined to be unobservable by the enhanced BDD if, in the absence of noise, there exists a state attack matrix $C \neq 0$ that satisfies \eqref{eq:uFDIattack} and 
	\begin{equation}
		CA^T=0.\label{eq:KCLofZIB}
	\end{equation}
	Recall that $\mathcal{J}$ is the subset of PMU measurements controlled by the attacker. This attack is feasible if and only if $supp(CH^T)\subseteq \mathcal{J}$. 
	
	Suppose the attacker wishes to change the state at a single bus. As we have shown through the example in Sec. \ref{EBDD}, it may not be possible to launch this attack without detection by just changing the state at this bus. What we wish to find out is the conditions under which this attack is feasible. The goal of this attack is to change the state at a single bus, but not necessarily changing the state only at this bus.
	
    Similar to \cite{Hug2012}, let $b$ be the bus that the attacker wishes to change state, the attack subgraph $\mathcal{S}_b$ can be constructed as follows.
	\begin{itemize}
		\item[1.] Include bus $b$ in $\mathcal{S}_b$.
		\item[2.] Include all buses adjacent to bus $b$ in $\mathcal{S}_b$.
		\item[3.] If there is a ZIB on the boundary of $\mathcal{S}_b$, extend $\mathcal{S}_b$ to include all adjacent buses of such a ZIB.
		\item[4.] Repeat step 3 until no bus on the boundary is ZIB.
	\end{itemize}
	When more than one bus are to be attacked, the final subgraph will be given by the union of all these subgraphs. 
	\begin{equation}
		\mathcal{S}=\cup \mathcal{S}_b, \forall b \in supp(C). \label{eq:Subgraph}
	\end{equation}
	
\begin{figure}[]
		\hspace{-0cm}\centering \includegraphics[trim=0 0.3cm 0 0.3cm, scale=0.85]{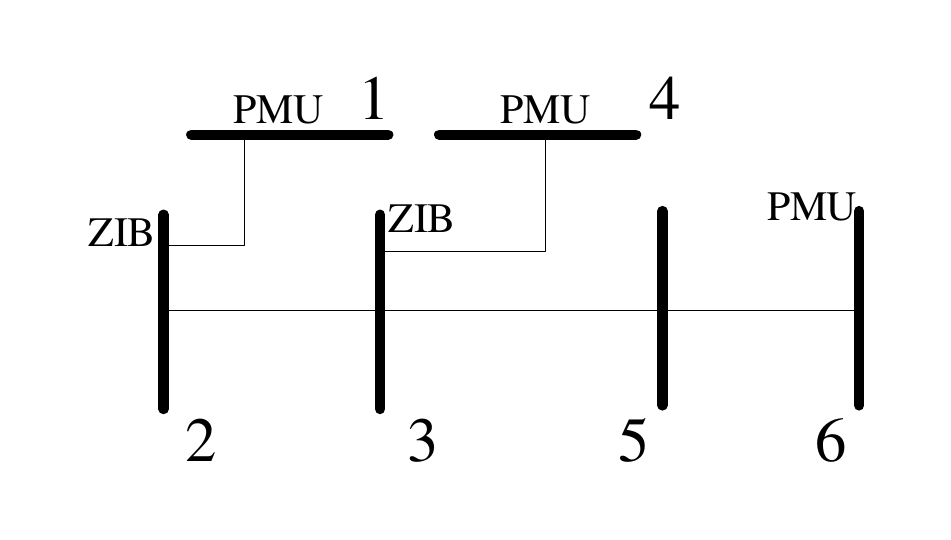}\protect\protect\caption{6-bus system illustrating that coltrolling all PMUs in $\mathcal{S}$ is unnecessary to launch unobservable FDI attacks \label{SixBus}}
	\end{figure}

	An attacker who has control over all PMUs in $\mathcal{S}$ can launch an unobservable FDI attack. If there are no ZIBs in $\mathcal{S}$, this condition is sufficient and necessary to make this attack unobservable. However, this is merely a sufficient condition, if there are ZIBs in $\mathcal{S}$. We now illustrate an example through the 6-bus system in Fig. \ref{SixBus}. Suppose the attacker wishes to change the state at bus $5$. If we apply the subgraph approach described above, the resulting subgraph is the whole system. Thus, it would appear that the attacker needs to control all three PMUs. However, if the attacker controls the PMUs at bus $4$ and $6$, it can launch this attack without being detected. If the attacker wants to change $V_5$, $I_{5,6}$ and $I_{3,5}$ are changed. Since the attacker does not control the PMU at bus $1$, then according to KCL of ZIB $2$, $V_1,V_2,I_{1,2},I_{2,3}$ cannot be changed, and consequently $V_3$ cannot be changed. To satisfy KCL at bus $3$, the attacker can change $I_{3,4}$ and $V_4$, and then this attack becomes feasible. Clearly, this attack is feasible under the condition that attacker only control PMUs at bus $4$ and $6$. 
    Generalizing these conditions is out of the scope of this paper.
	
	\section{Multiplicative Attacks}\label{sec:RM-FDI}
    \subsection{Prior Work: Attack Detection Based on Low-Rank Matrix Decomposition} \label{sec:LRD}
    Conventional BDDs cannot detect unobservable FDI attacks in the form of \eqref{eq:FDIattack}. However, exploiting the low-rank nature of the high-dimensional PMU data matrix $W$, the authors in \cite{Wang_SmartGridComm2014_PMU} propose a new attack detection mechanism based on LRD so as to separate the low-rank matrix $W$ and column sparse matrix $CH^T$ in \eqref{eq:FDIattack}. The LRD detector is briefly reviewed below. 

    Given a post-attack measurement matrix $\bar{W}$, an estimate $\hat{W}$ of the actual measurement matrix, and the attack matrix $\hat{C}$ are formed by solving the following convex optimization problem:   
    \begin{subequations}\label{LRD_Detector}
	\begin{flalign}
	    \underset{\hat{W}\in\mathbb{C}^{N\times n},\hat{C}\in \mathbb{C}^{N\times p}}{\text{minimize}} \;\;& \|\hat{W}\|_* + \lambda \|\hat{C}\|_{1,2} \label{eq:LRD_OBJ}\\
	    \text{subject to} \hspace{0.6cm}& \bar{W}=\hat{W}+\hat{C}\bar{H}^T \label{eq:LRD_UnobservableAttack}
	\end{flalign}
    \end{subequations} 
    where $\|\hat{W}\|_*$ is the nuclear norm of $\hat{W}$; $\|\hat{C}\|_{1,2}$ is the $l_{1,2}$-norms of $\hat{C}$, \textit{i.e.,} the sum of $l_2$-norm of columns in $\hat{C}$; $\lambda$ is a weighting parameter; and $\bar{H}$ is the normalized Jacobian matrix, where for each row vector $H_i$, $\bar{H}_i=H_i / \lVert H_i\rVert$.  The objective \eqref{eq:LRD_OBJ} is to minimize the rank of $\hat{W}$ (captured by its nuclear norm) and the column sparsity of $\hat{C}$ (captured by its $l_{1,2}$-norm). 

    After obtaining the optimal solution, $(\hat{W}^*,\hat{C}^*)$ for \eqref{LRD_Detector}, the set of attacked measurements and states can be identified as $supp(\hat{C}^*\bar{H}^T)$ and $supp(\hat{C}^*)$, respectively. Assume there exists unobservable attacks in $\bar{W}$, such that $\bar{W}=W+C\bar{H}^T$. The authors of \cite{Wang_SmartGridComm2014_PMU} prove that for a specific range of $\lambda$, \textit{i.e.,} $\lambda \in \left[\lambda_\text{min},\;\lambda_\text{max}\right]$, the optimization in \eqref{LRD_Detector} can successfully identify $supp(C)$, \textit{i.e.,} $supp(\hat{C}^*)=supp(C)$, under the assumption that every non-zero column of $C\bar{H}^T$ does not lie in the column space of $W$.
    
    \subsection{Attack Design}
    We assume that the attacker has the following knowledge and capabilities:
    	\begin{enumerate}
    		\item The attacker has full system topology information.
    		\item The attacker has control of the measurements in $\mathcal{J}$.
    	\end{enumerate}
	Given a PMU measurement vector $w_t$ at time $t$, an attacker can estimated state vector $\hat{x}_t$ using \eqref{eq:LinearSE}, and construct false measurements as
		\begin{flalign}
		\bar{w}_t = H^T F^T \hat{x}_t,\label{eq:falseMeasurements}
		\end{flalign}
		where $F$ is a full rank matrix.
    If the attacker does this for $N$ time instances, the resulting PMU measurement matrix can be written in the form of \eqref{eq:uFDIattack} as:
    \begin{flalign}
    	\bar W = \hat{X}FH^T=&W + CH^T = \hat{X}H^T + CH^T,\\
    	 &C = X(F-I).
    \end{flalign}
    In order to remain unobservable, the attack must satisfy \eqref{eq:KCLofZIB}:
    \begin{flalign}
    	CA^T = &\hat{X}(F-I)A^T=XFA^T-XA^T\notag\\
    	     = & \hat{X}FA^T=0. \label{eq:uAttackF}
    \end{flalign}
    This class of attacks do not change the rank of the measurement matrix $W$, and is potentially undetectable by the LRD detector. However, the nuclear norm of $W$ may change under attack. Since the LRD detector uses the nuclear norm of the measurement matrix as a proxy of its rank, these attacks may still be detected by the LRD detector.
	\section{Numerical Results} \label{sec:Simulation}
	In this section, we illustrate the efficacy of the unobservable FDI attacks introduced in Sec. \ref{sec:RM-FDI}. To this end, we construct the post-attack measurement matrix $\bar{W}$ with as in \eqref{eq:falseMeasurements}. Subsequently, we solve the LRD detection optimization problem \eqref{LRD_Detector} for $\bar{W}$ to check if the attack matrix $C$ is detected. We assume that the LRD detector selects 5 seconds worth of PMU measurements data, \textit{i.e.,} $N=150$, while the attacker continuously injects bad data. The test system is the IEEE RTS-24-bus system. The convex optimization problems for LRD detection is solved with MOSEK on a 3.40 GHz PC with 32 GB RAM.  $\lambda$ is chosen to be 1.05 in the LRD detector.

    \subsection{Synthetic PMU Measurements Generation}
\begin{figure}[h]
	\centering{}\includegraphics[trim=0 1cm 0 1cm,scale=0.45]{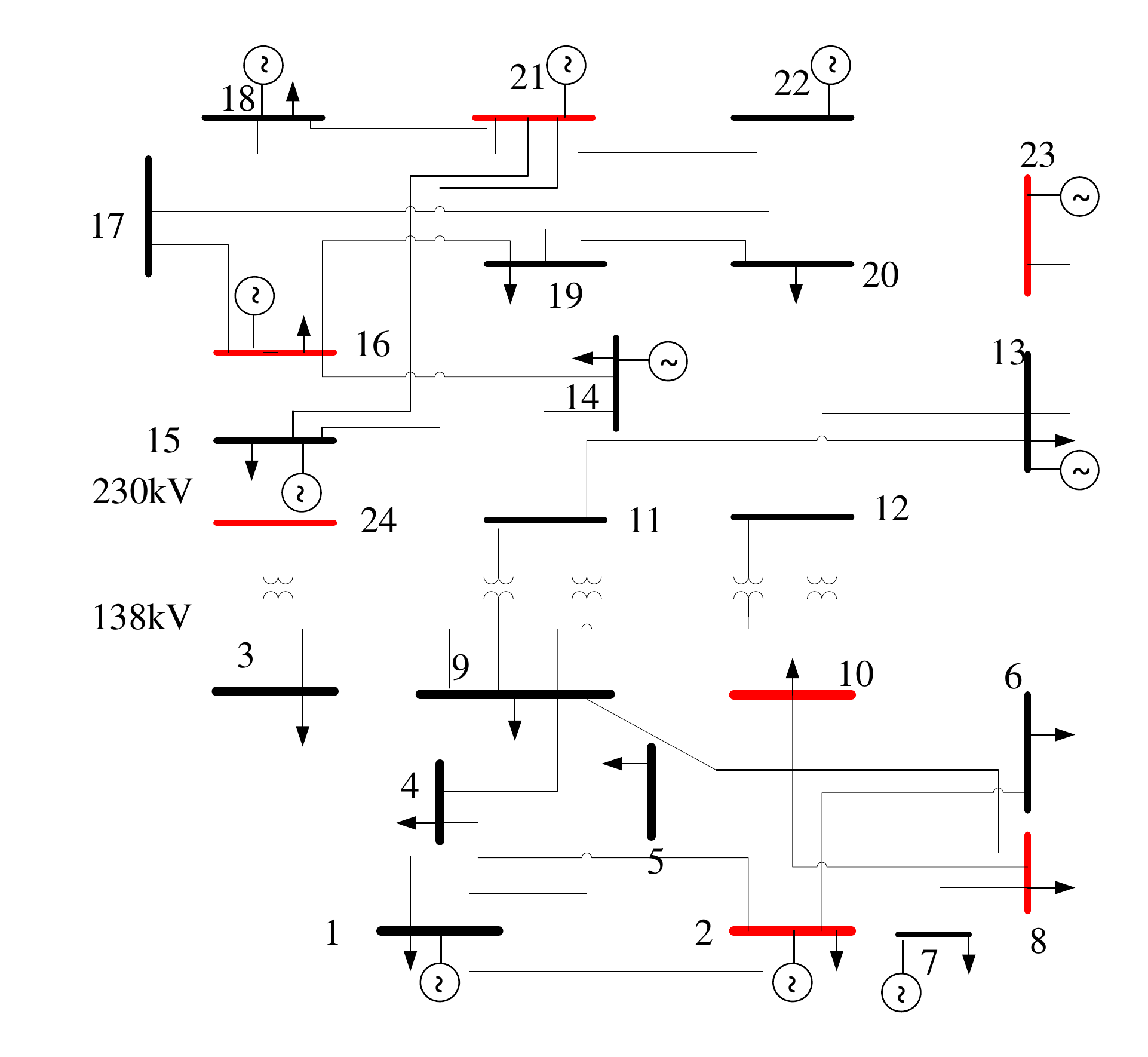}\protect\protect	\caption{PMU placement in the IEEE RTS-24-bus system.\label{fig:OPP}}
\end{figure}

\begin{figure}[h]
		\centering{}\includegraphics[trim=0 1cm 0 1cm,scale=0.45]{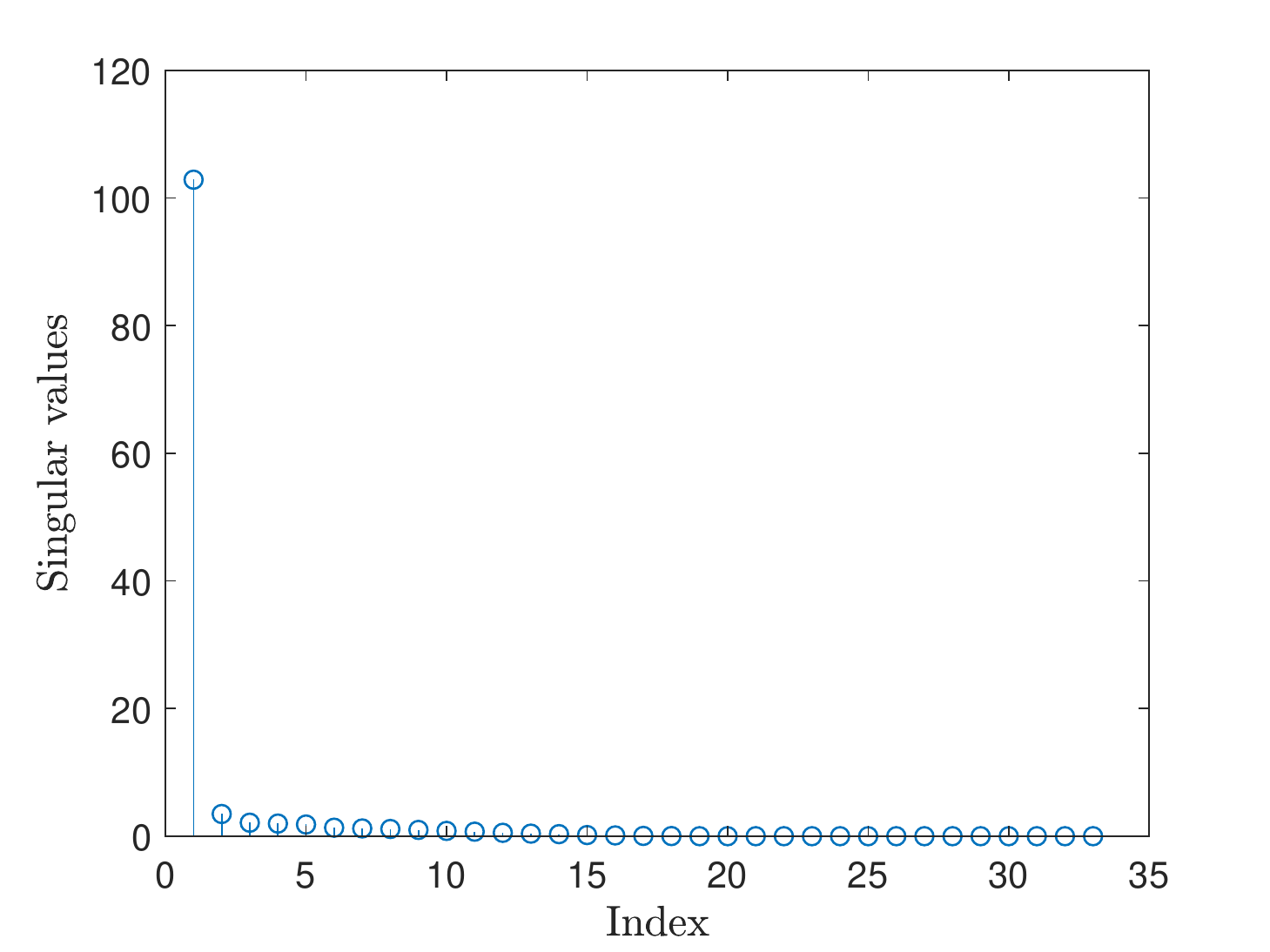}\protect\protect	\caption{Singular values of the synthetic PMU data matrix in decreasing order.\label{fig:SingularValue}}
	\end{figure}

	An optimal PMU placement problem as introduced in \cite{PMUplacement2004} is solved to ensure the system fully observable with PMUs. The details of the PMU placement scheme and available measurements are illustrated in Fig. \ref{fig:OPP}. Buses in red are buses with PMUs. We generate synthetic PMU data over 5 seconds in the test system. A base case of the system operating status is obtained by solving an AC optimal power flow problem. To model realistic data with a disturbance, at the first time instant $t$ after 1 second, we change the load at each bus by adding a random value $d$ to the base load, such that $d\sim \mathcal{N}\left(0,\frac{60}{1.1^{\left(t-31\right)}}\right)$. We then solve an AC power flow to obtain the measured phasors of bus voltage and branch current as measurements at time instant $t$. The singular values for the synthetic measurement matrices are illustrated in Fig \ref{fig:SingularValue}. It can be seen that these synthetic measurements have the same low-rank property as the actual PMU data as illustrated in \cite{Wang_SmartGridComm2014_PMU}.  Furthermore, we assume noiseless measurements, \textit{i.e.}, $E=\mathbf{0}$.

	\subsection{Detection of Multiplicative FDI Attacks}
	We focus on unobservable 1-sparse multiplicative attacks and exhaustively test their detectability. Due to space limitations, we illustrate our results using two specific cases. One case illustrates that no attacks are detected, and the other case illustrates that attacks are detected at incorrect buses. To change the state at bus $b$, the entries of the attack matrix $F$ are set to be
	\begin{equation}\label{Fsetup}
		F_{ij} = \left\{ \begin{array}{lr}
		1, \hspace{0.5cm}i=j\neq b \\
		c,  \hspace{0.5cm}i=j=b\\
		0,  \hspace{0.5cm}i\neq j.
		\end{array}\right.
	\end{equation}
	Clearly $F$ is a diagonal matrix, and hence $F$ is full rank. Note that due to the effects of ZIBs in \eqref{eq:uAttackF}, it is not feasible to launch 1-sparse unobservable attacks targeting some buses, since changing their states requires the attacker to also change the state of other buses, as illustrated in the example in Sec. \ref{FeasCondition}. 
	
\begin{figure}[h]
		\centering{}\includegraphics[trim=0 1cm 0 1cm,scale=0.6]{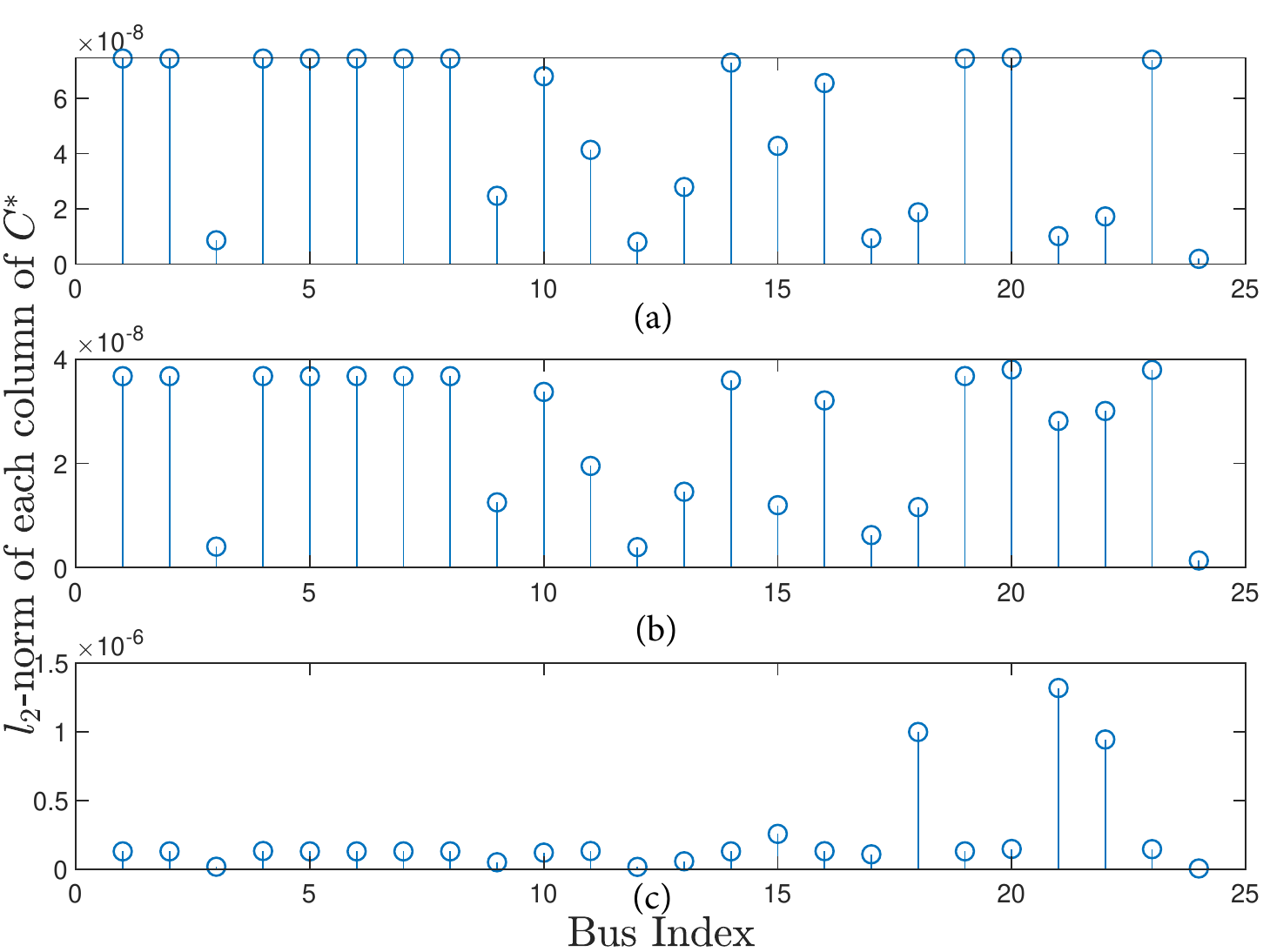}\protect\protect
		\vspace{0.3cm}	\caption{Normalized $l_2$-norm of each column of $\hat{C}^*$ under (a) no attack; (b) attack at bus 4; (c) attack at bus 16\label{fig:Bus4Atk}}
	\end{figure}

	Fig. \ref{fig:Bus4Atk}(a) illustrates the detection result when there is no attack.
	Fig. \ref{fig:Bus4Atk}(b) illustrates the attack detection result for an attack on bus $4$. $F$ is constructed as in \eqref{Fsetup} with $b=4$ and $c=e^{j0.2}$. Compare these two subfigures, we conclude that the LRD detector fails to detect the attack at bus $4$.
	Fig. \ref{fig:Bus4Atk}(c) illustrates the attack detection result for an attack on bus $16$. $F$ is constructed as in \eqref{Fsetup} with $b=16$ and $c=e^{j0.3}$. The LRD detector incorrectly detects that buses $18$, $21$, and $22$ are under attack. 
	\vspace{0.3cm}
	
\begin{figure}[h]
			\centering{}\includegraphics[trim=0 1cm 0 1cm,scale=0.6]{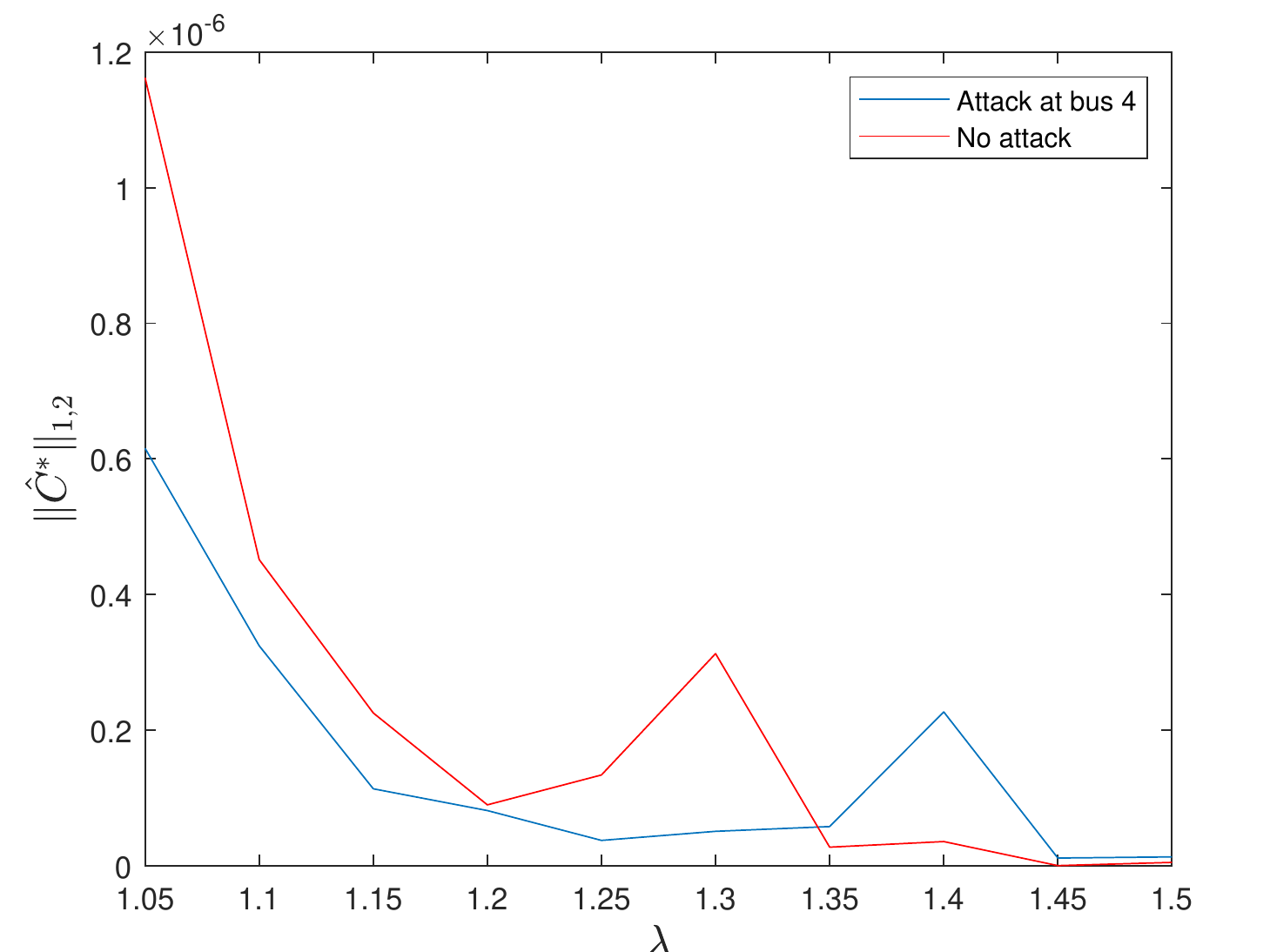}\protect\protect
			\vspace{0.3cm}	\caption{$l_{1,2}$-norm of $\hat{C}^*$ under no attack and attack at bus 4 for different $\lambda$\label{fig:comp}}
		\end{figure}

	A comparison of the $l_{1,2}$-norm of $\hat{C}^*$ with no attack and with the attack at bus $4$ for $\lambda=[1.05,1.5]$ is illustrated in Fig. \ref{fig:comp}. It can be seen that the $\| \hat{C}^*\|_{1,2}$ with and without this attack are very similar. Intuitively, this attack cannot be detected by the LRD detector, and our result in Fig. \ref{fig:Bus4Atk}(b) supports such intuition.

	\section{Concluding Remarks}
	In this paper, unobservable FDI attacks against PMUs have been studied. We have introduced an enhanced BDD, and have given conditions under which the attacks are unobservable. Prior work demonstrated that unobservable FDI attacks can be identified by the LRD detector. In this work, we have introduced a class of multiplicative attacks that can potentially bypass the LRD detector. Future work involves generalizing the conditions for launching unobservable FDI attacks and developing countermeasures for such attacks.
	
	\vspace{0.3cm}

	\section*{Acknowledgment}
	This material is based upon work supported by the National Science Foundation under Grant No. CNS-1449080.
	

	%
	%

	
	
	%
	%
	%
	
	\bibliographystyle{IEEEtran}
	\bibliography{dis}

\end{document}